\documentclass[runningheads]{llncs}

\usepackage{eccv}

\usepackage{eccvabbrv}
\usepackage{hyperref}
\usepackage{url}
\usepackage{threeparttable} 

\usepackage{amsmath,amssymb,amsfonts}
\usepackage{algorithmic}
\usepackage{graphicx}
\usepackage{textcomp}
\usepackage{amsmath}
\usepackage{siunitx}
\usepackage{xcolor}
\usepackage{multirow}
\usepackage{wrapfig}
\usepackage{algorithm}
\usepackage{pifont}
\usepackage{enumitem}
\usepackage{lmodern,babel,adjustbox,booktabs,multirow}

\usepackage{wrapfig}

\usepackage{graphicx}
\newcolumntype{L}[1]{>{\raggedright\arraybackslash}p{#1}}
\newcolumntype{C}[1]{>{\centering\arraybackslash}p{#1}}
\newcolumntype{R}[1]{>{\raggedleft\arraybackslash}p{#1}}
\newcommand*\samethanks[1][\value{footnote}]{\footnotemark[#1]}

\usepackage[accsupp]{axessibility}  

\usepackage{hyperref}

\usepackage{orcidlink}

\begin{document}

\title{FairDomain: Achieving Fairness in Cross-Domain Medical Image Segmentation and Classification} 

\titlerunning{FairDomain}

\author{Yu Tian\inst{1\thanks{Contributed equally as co-first authors.}}\orcidlink{0000-0001-5533-7506} \and
Congcong Wen\inst{2,3\samethanks}\orcidlink{0000-0001-6448-003X} Min Shi\inst{1\samethanks}\orcidlink{0000-0002-7200-1702}  \and  Muhammad Muneeb Afzal\inst{3}  \and \\ Hao Huang\inst{2,3}\orcidlink{0000−0002−9131−5854} \and Muhammad Osama Khan\inst{3}\orcidlink{0009-0001-0897-3283}
 \and Yan Luo\inst{1}\orcidlink{0000-0001-5135-0316}  \and \\  Yi Fang\inst{2,3\thanks{Contributed equally as co-senior authors.}}\orcidlink{0000-0001-9427-3883} \and Mengyu Wang\inst{1\samethanks}\orcidlink{0000-0002-7188-7126}
}

\authorrunning{Y.~Tian, C.~Wen, M.~Shi, et al.}

\institute{Harvard Ophthalmology AI Lab, Harvard University \\ \and
Center for Artificial Intelligence and Robotics, New York University Abu Dhabi\\ \and Embodied AI and Robotics (AIR) Lab, New York University
}

\maketitle
\vspace{-15pt}

\begin{abstract}

Addressing fairness in artificial intelligence (AI), particularly in medical AI, is crucial for ensuring equitable healthcare outcomes. Recent efforts to enhance fairness have introduced new methodologies and datasets in medical AI. However, the fairness issue under the setting of domain transfer is almost unexplored, while it is common that clinics rely on different imaging technologies (e.g., different retinal imaging modalities) for patient diagnosis. This paper presents FairDomain, a pioneering systemic study into algorithmic fairness under domain shifts, employing state-of-the-art domain adaptation (DA) and generalization (DG) algorithms for both medical segmentation and classification tasks to understand how biases are transferred between different domains. We also introduce a novel plug-and-play fair identity attention (FIA) module that adapts to various DA and DG algorithms to improve fairness by using self-attention to adjust feature importance based on demographic attributes. Additionally, we curate the first fairness-focused dataset with two paired imaging modalities for the same patient cohort on medical segmentation and classification tasks, to rigorously assess fairness in domain-shift scenarios. Excluding the confounding impact of demographic distribution variation between source and target domains will allow clearer quantification of the performance of domain transfer models. Our extensive evaluations reveal that the proposed FIA significantly enhances both model performance accounted for fairness across all domain shift settings (i.e., DA and DG) with respect to different demographics, which outperforms existing methods on both segmentation and classification. The code and data can be accessed at \url{https://ophai.hms.harvard.edu/datasets/harvard-fairdomain20k}.
  \keywords{Fairness Learning \and Domain Shift \and Medical Segmentation \and Medical Classification}
\end{abstract}

\section{Introduction}
\label{sec:intro}

Advancements in deep learning have revolutionized the field of medical imaging, enabling significant improvements in tasks such as classification~\cite{tian2023self,tian2023unsupervised,liu2022acpl,shi2024rnflt2vec,tian2021constrained,shi2023artifact,liu2022nvum,chen2022multi,wang2023learning,chen2023bomd,chen2024braixdet} and segmentation~\cite{liu2022translation,chen2021transunet,zhou2018unet++,wang2022medical,tian2022contrastive,li2014medical,liu2022translation}. These technologies have the potential to enhance diagnostic accuracy, streamline treatment planning, and ultimately improve patient outcomes. Despite these advancements, the deployment of deep learning models across varied healthcare settings has unearthed a pivotal challenge: the risk of inherent algorithmic bias and discrimination against certain demographic groups, which could undermine the fairness of medical diagnostics and treatments.

Recent studies have begun to address the issues of algorithmic biases in medical imaging by developing methodologies aimed at enhancing the fairness of deep learning models~\cite{tian2024fairseg,luo2024fairvisionequitabledeeplearning,luo2023harvard,dressel2018accuracy,shi2024equitable,luo2024fairclip,asuncion2007uci,wightman1998lsac,miao2010did,kuzilek2017open,ruggles2015ipums,zhang2017age,zong2022medfair,irvin2019chexpert,johnson2019mimic,kovalyk2022papila}. These methodologies, while pioneering, commonly presuppose that the distribution of data during training and testing remains constant, thereby assuming that fairness measures implemented during training will suffice to ensure equitable decisions during testing within identical domains. This presumption, however, frequently does not hold in practical healthcare scenarios. For instance, primary care clinics and specialty hospitals may rely on different imaging technologies (e.g., different retinal imaging modalities) for patient diagnosis, leading to significant domain shifts that can adversely affect model performance and fairness when models trained on one type of imaging data are deployed on another. 
Therefore, it is critical to account for domain shifts and learn fair models that are robust to potential cross-domain scenarios in real-world deployment environments. 

\begin{figure*}[t!]
    \centering
    \includegraphics[width=1\textwidth]{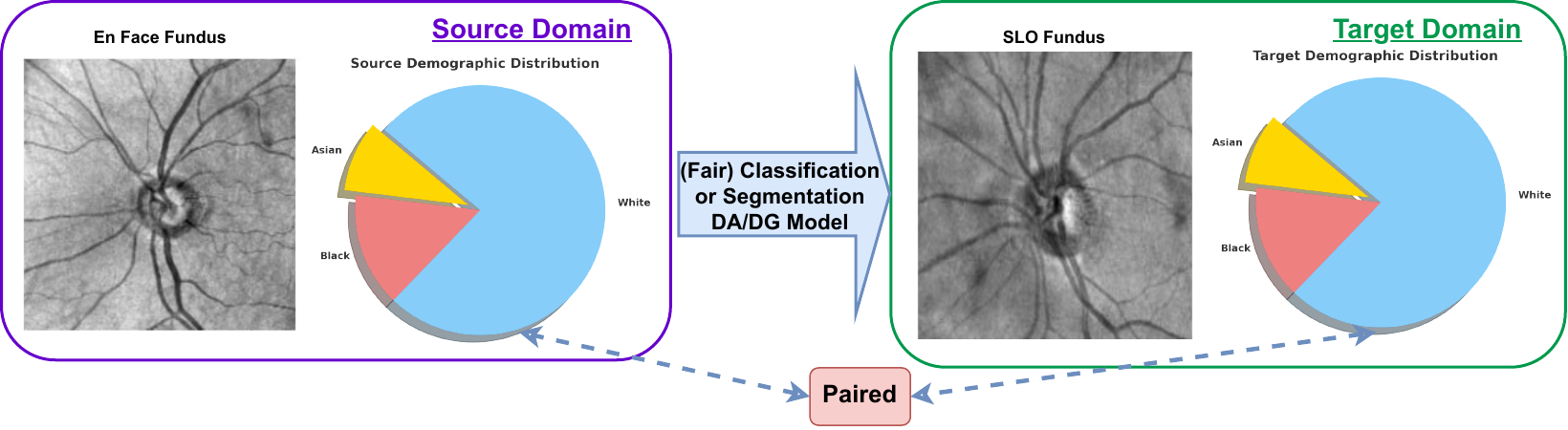}
    \vspace{-12pt}
    \captionof{figure}{\textbf{FairDomain overview}. 
    An example of domain adaptation and generalization in healthcare: (fair) segmentation or classification model trained with patient data with one imaging modality, such as En Face Fundus images, can be adapted or generalized to another imaging modality such as SLO Fundus images, by maintaining both high accuracy and fairness.
    }
    \label{fig:intro}
\end{figure*}

The previous literature extensively explores domain adaptation and domain generalization as methodologies to counteract the challenges posed by domain shifts, aiming to develop models that perform reliably across diverse but related domains. Domain adaptation, particularly in its unsupervised form, leverages both labeled data from a source domain and unlabeled data from a target domain to facilitate model generalization to new, unseen data~\cite{zhao2018adversarial}. Conversely, domain generalization operates under the premise that target domain data remain inaccessible during training, relying solely on source domain data to foster model applicability to novel, unseen domains~\cite{qiao2020learning}. Despite extensive research into enhancing model \textbf{accuracy} across domain shifts, the critical aspect of \textbf{fairness} — ensuring that models provide equitable predictions across different demographic groups — has been notably underexplored. This insufficiency is particularly critical in medical domains, where decision-making models directly impact human health, well-being, and safety. Only limited studies have begun to explore the transfer of fairness across domains~\cite{obermeyer2021algorithmic,pham2023fairness,mukherjee2022domain}. 
However, all these previous studies lack systematically comprehensive investigations of fairness issues under domain shift.
They predominantly focus on either domain adaptation or generalization, but rarely both. Furthermore, existing studies primarily address medical classification challenges, overlooking the critical task of medical segmentation, which is equally if not more significant under domain shifts in healthcare scenarios. Additionally, many studies adopt impractical assumptions about domain shifts, such as considering changes in patient age groups as domain shifts, which contradicts established understandings of medical domain shifts that are typically caused by the use of different imaging technologies~\cite{guan2021domain, li2020domain}.

\begin{figure}[t]
\centering
    \begin{subfigure}[b]{0.3\textwidth}
            \centering
            \includegraphics[width=1\linewidth]
    {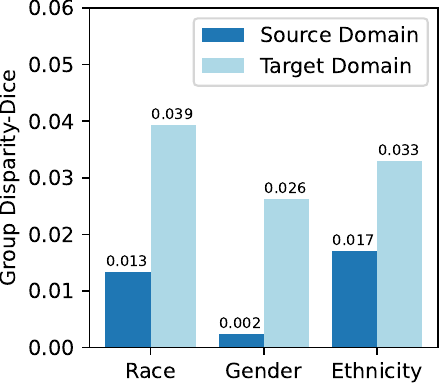}
    \caption{Cup segmentation}
    \label{fig:summarization_methods}
    \end{subfigure} \hspace{2px}
    \begin{subfigure}[b]{0.3\textwidth}
            \centering
            \includegraphics[width=1\textwidth]{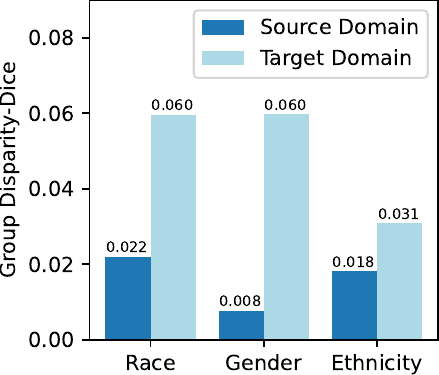}
        \caption{Rim segmentation}
    \label{fig:ablation_pmcclip}
    \end{subfigure} \hspace{2px}
    \begin{subfigure}[b]{0.3\textwidth}
            \centering
            \includegraphics[width=1\textwidth]{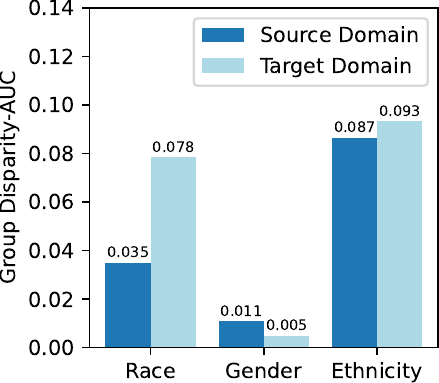}
        \caption{Glaucoma classification}
    \label{fig:ablation_advloss}
    \end{subfigure} 
    \caption{Group performance disparities in the source and target domains. It is evident that as domain shift occurs, algorithmic fairness significantly deteriorates throughout this process.}
    \label{fig:gd}
    \vspace{-20pt}
\end{figure}

To address these gaps, our work introduces the first comprehensive systematic investigation of algorithmic fairness under domain shift in the medical imaging field, named FairDomain. We conduct extensive experiments across multiple state-of-the-art (SOTA) domain adaptation and generalization algorithms with three common demographic attributes, to assess both the accuracy and fairness of those algorithms and understand how fairness transfers across different domains in terms of both medical segmentation and classification tasks, as shown in Fig.~\ref{fig:intro}. In Fig.~\ref{fig:gd}, we investigate the transfer of algorithmic fairness across domains. Our observations reveal a significant exacerbation of group performance disparities between source and target domains (i.e., increased disparity performance values) when subjected to domain shifts in different medical classification and segmentation tasks. This indicates the critical necessity for devising fairness-oriented algorithms to address this pressing issue effectively.
Recognizing the limitations of previous bias mitigation efforts, which were confined to single domains and lacked adaptability to diverse domain shift scenarios, we introduce a novel, versatile fair identity attention (FIA) mechanism. This technique, designed for seamless integration with various domain adaptation and generalization strategies, utilizes self-attention derived from demographic attributes (e.g., racial groups) to harmonize feature significance across demographics, which will promote fairness.
A pivotal challenge in developing the FairDomain benchmark was the absence of a medical imaging dataset that realistically encapsulates domain shifts in real-world medical domains, where domain shifts are typically induced by disparate imaging technologies, without conflating with demographic distribution shifts. Existing medical datasets with differences in patient demographics between source and target domains introduce ambiguity regarding the root causes of observed algorithmic bias — is it the performance fairness change in domain transfer due to demographic distributions or the inherent domain shift? To address this, we curated a unique dataset with paired retinal fundus images from the same patient cohort, captured through two distinct imaging modalities (En face and SLO fundus images), specifically for analyzing algorithmic bias in domain-shifted scenarios. To our knowledge, this is the first dataset meticulously designed to study fairness in domain shifts with consistent patient distributions, thus offering a clearer analysis of algorithmic biases attributable solely to domain shifts.
Compared with previous medical datasets, our FairDomain dataset is distinguished by three key features: (1) It includes cross-domain data for both medical \textbf{segmentation} and \textbf{classification} tasks, offering a holistic view of algorithmic fairness under domain shifts; (2) The pairing of source and target domain images for the same patient cohort allows a clean-cut setting for studying the model fairness change due to domain shift excluding the confounding impact of demographic distribution variation on model fairness change; and (3) The dataset is enriched with six real-world demographic attributes from large eye hospitals, facilitating a more nuanced study of fairness. This new dataset serves as an invaluable resource not only for fairness research in cross-domain tasks but also for general cross-domain tasks as excluding the confounding impact of demographic distribution variation between source and target domains will allow clearer quantification of the performance of domain transfer models.

To summarize, our contributions are threefold:
\begin{itemize}
    \item The first systematic exploration of algorithmic fairness under domain shifts in medical imaging.
     \item  The introduction of fair identity attention techniques to improve accuracy and fairness across domain adaptations and generalizations.
     \item The creation of a large-scale, paired medical segmentation and classification dataset for fairness studies under domain shifts. 
\end{itemize}
 

\section{Related Work}
\textbf{Domain Adaptation and Generalization:} Unsupervised Domain Adaptation (UDA) leverages unlabeled target data has become essential, with adversarial learning~\cite{hoffman2016fcns,tsai2018learning,vu2019advent,chen2018road} and self-supervised training~\cite{zou2018DAseg,mei2020instance} standing out as principal strategies to harmonize feature distributions across different domains. These methods, complemented by innovations like pseudo labeling~\cite{chen2017no}, depth-aware adaptation~\cite{vu2019dada}, and entropy minimization~\cite{vu2018advent}, have significantly advanced UDA. In parallel, Domain Generalization (DG) seeks to cultivate domain-invariant features through adversarial techniques~\cite{ganin2016domain,li2018deep}, regularization~\cite{motiian2017unified,ghifary2015domain}, instance reweighting~\cite{matsuura2020domain}, and meta-learning~\cite{li2018learning,balaji2018metareg}. Recent shifts towards style-based learning, which distinguishes between the content and style of images to mitigate domain gaps~\cite{pan2018two,choi2020hi,yang2020fda,wang2020learning,xu2021fourier,zhao2021test}. Nevertheless, these advancements primarily focus on enhancing model accuracy and overlook incorporating fairness into adaptation and generalization frameworks.

\noindent\textbf{Fairness Learning in Medical Imaging:} The pursuit of fairness in machine learning, particularly in medical imaging, aims to mitigate biases and ensure equitable outcomes across diverse patient groups. Recent works have introduced fairness datasets~\cite{luo2023harvard,tian2024fairseg,luotmi,irvin2019chexpert,johnson2019mimic,groh2021evaluating} and methodologies~\cite{quadrianto2019discovering,ramaswamy2021fair,zhang2020towards,park2022fair,beutel2017data,roh2020fr,sarhan2020fairness,zafar2017fairness,zhang2018mitigating,wang2022fairness,kim2019multiaccuracy,lohia2019bias} to study and improve algorithmic fairness. However, these efforts often overlook the dynamic nature of healthcare settings, where various medical imaging modalities may used from different healthcare domains, and therefore model fairness issues should be addressed in the context of cross-domain modeling.

\noindent\textbf{Bridging Fairness under Domain Shift:} The intersection of fairness under domain shift represents an emerging research frontier. Only a few existing works study the transfer of algorithmic bias under domain shift~\cite{pham2023fairness,truong2023fredom,zhang2021aligning,pham2023fairness}.  Pham et al.~\cite{pham2023fairness} focus on proposing a theoretical framework to understand how fairness and accuracy transfer by density matching and Truong~\cite{truong2023fredom} exemplify efforts to integrate fairness into domain adaptation processes, aiming to ensure equitable predictions across class distributions. Additionally, Zhang et al.~\cite{zhang2021aligning} focus on using domain adaptation approaches to improve individual fairness. However, these studies predominantly focus on the theoretical frameworks of domain shift on limited tasks and lack comprehensive systemic empirical studies of fairness issues under domain shift. Moreover, prior
methodologies often rely on restrictive assumptions regarding domain shifts due to a lack of high-quality domain shift datasets for a better study of this problem. The existing datasets and settings might not accurately reflect real-world scenarios, particularly in the context of medical imaging settings.

\section{Dataset Analysis}
\label{section:data}


\noindent\textbf{Data Collection and Quality Control.}
Our institute's institutional review board (IRB) approved this study, which followed the principles of the Declaration of Helsinki. Since the study was retrospective, the IRB waived the requirement for informed consent from patients.

The subjects tested between 2010 and 2021 are from a large academic eye hospital of Harvard Medical School. Two cross-domain tasks, namely medical segmentation and classification, are studied in this work. For medical segmentation, there are five types of data: (1) En-face fundus imaging scans; (2) SLO fundus imaging scans; (3) patient demographics; (4) glaucoma diagnosis; and (5) cup-disc masks. Particularly, the pixel annotations of cup and disc regions are first acquired from the OCT machine, where the disc border in 3D OCT is segmented as the Bruch's membrane opening by the OCT manufacturer software, and the cup border is detected as the intersection between the inner limiting membrane (ILM) and the plane that results in minimum surface area from the intersection and disc border on the plane~\cite{mitsch2019comparison, everett2015automated}. Approximately, the cup border can be considered as the closest location on the ILM to the disc border, which is defined as the Bruch's membrane opening. Both Bruch's membrane opening and the internal limiting membrane can be easily segmented due to the high contrast between them and the background. Since the OCT manufacturer software leverages 3D information, the cup and disc segmentation is generally reliable.  
Given the limited availability and high cost of OCT machines in primary care, we propose a method to transfer annotations from 3D OCT to 2D SLO fundus images, aiming to enhance early-stage glaucoma screening. By utilizing the registration tool NiftyReg~\cite{modat2014global}, we accurately align SLO fundus images with OCT-derived pixel-wise annotations, generating a vast set of high-quality SLO fundus mask annotations. This process, verified by a panel of medical experts, shows an 80\% success rate in registrations, streamlining the annotation process for broader applications in primary care settings. Upon the alignment and manual examination of those annotations, we leverage pixel-wise masks from both the SLO and En face fundus images to examine the transfer of algorithmic fairness in segmentation models under domain shifts.

For medical classification, there are four types of data: (1) En-face fundus imaging scans; (2) SLO fundus imaging scans; (3) patient demographics; and (4) glaucoma diagnosis. The subjects in the medical classification dataset are categorized into two classes including normal and glaucoma defined based on visual field tests.

\noindent\textbf{Data Characteristics.}
The medical segmentation dataset contains 10,000 samples from 10,000 subjects. We divide our data into the training set with 8,000 samples, and the test set with 2,000 samples.
The patient age average is 60.3 $\pm$ 16.5 years. Within this dataset, we have six demographic attributes including age, gender, race, ethnicity, preferred language, and marital status. The demogra-phic distributions are as follows: \textbf{Gender:} Female: 58.5\%, and Male: 41.5\%; \textbf{Race:} Asian: 9.2\%, Black: 14.7\%, and White: 76.1\%. \textbf{Ethnicity:} Non-Hispanic: 90.6\%, Hispanics: 3.7\%, and Unknown: 5.7\%. \textbf{Preferred Language:} English: 92.4\%, Spanish: 1.5\%, Others: 1\%, and Unknown: 5.1\%. \textbf{Marital Status:} Married or Partnered: 57.7\%, Single: 27.1\%, Divorced: 6.8\%, Legally Separated: 0.8\%, Windowed: 5.2\%, and Unknown: 2.4\%.

Similarly, the medical classification dataset contains 10,000 samples from 10,000 subjects with an average age of 60.9 $\pm$ 16.1 years. We divide our data into the training set with 8,000 samples, and the test set with 2,000 samples. 
Within this dataset, we have six demographic attributes including age, gender, race, ethnicity, preferred language, and marital status. The demographic distributions are as follows: \textbf{Gender:} Female: 72.5\%, and Male: 27.5\%; \textbf{Race:} Asian: 8.7\%, Black: 14.5\%, and White: 76.8\%. \textbf{Ethnicity:} Non-Hispanic: 96.0\%, Hispanics: 4.0\%. \textbf{Preferred Language:} English: 92.6\%, Spanish: 1.7\%, Others: 3.6\%, and Unknown: 2.1\%. \textbf{Marital Status:} Married or Partnered: 58.5\%, Singe: 26.1\%, Divorced: 6.9\%, Legally Separated: 0.8\%, Windowed: 1.9\%, and Unknown: 5.8\%.


%

\section{Method}

\subsection{Problem Statement}

Domain Adaptation (DA) and Domain Generalization (DG) are pivotal techniques in the development of machine learning models, aimed at addressing the variability that can occur when a model trained in one specific domain is applied to another. In the field of medical imaging, the techniques of DA and DG are critical in creating models that can robustly handle the variability present across different medical institutions, imaging devices, and patient populations. In this paper, we aim to explore the dynamics of fairness within the context of domain shift and develop methodologies to ensure that models remain fair and reliable as they are adapted or generalized to new domains. The problem definitions of DA and DG are as follows:


 \noindent\textbf{Domain Adaptation (DA):} Given a domain data \( D_S=\{ (x_m^S, y_m^S, a_m^S) \}_{m=1}^{N_S} \), where \( x_m^S \in \mathbb{R}^{H \times W} \) represents the \( m \)-th image in the source domain. For image classification, \( y_m^S \in \mathbb{R}^K \) is the category label indicating the class to which the image belongs. For image segmentation, \( y_m^S \in \mathbb{R}^{K \times H \times W} \) is the label matrix with each element specifying the class for each pixel. In both contexts, \( a_m^S \in \mathbb{R}^{d} \) denotes an identity attribute associated with the patient, such as gender, race, or ethnicity. Additionally, there is an unlabeled target domain \( D_T = \{ x_m^T \}_{m=1}^{N_T} \), where the data distribution typically differs from the source domain. In addition to minimizing the distribution discrepancy between the source and target domains for effective performance on the target domain, a significant concern in DA is to ensure that the adaptation process respects fairness. This entails developing adaptation strategies that prevent the amplification of any biases present in the source domain and promote equity in treatment and outcomes across different groups defined by the identity attributes \( a_m^S \) in both the source and target domains.

 \noindent\textbf{Domain Generalization (DG):} Consider a setting with labeled source domain data \( D_S = \{ (x_m^S, y_m^S, a_m^S) \}_{m=1}^{N_S} \). Unlike DA, the DG approach does not utilize data from the target domain during training. Instead, the objective is to learn a model \( f_\theta \) from \( D_S \) that can generalize effectively to any related but unseen target domain \( T \), characterized by a potential shift in data distribution. The challenge in DG is to extract robust domain-invariant features from \( D_S \) that are predictive of the target domain \( T \) while also ensuring that the model's predictions are fair and unbiased with respect to the identity attributes represented by \( a_m^S \),  without having any access to \( T \) during training.


\subsection{Fair Identity Attention}

We aim to develop a methodical function \( f \) that mitigates fairness deterioration, which is often observed during the transfer of models from a source domain to a target domain. Such deterioration is primarily due to domain shift that can amplify existing biases in the dataset, notably those associated with demographic attributes like gender, race, or ethnicity. To address this, we propose an attention 
\begin{wrapfigure}{r!}{0.36\textwidth}
  \centering
  \vspace{-15pt}  \includegraphics[width=0.36\textwidth]{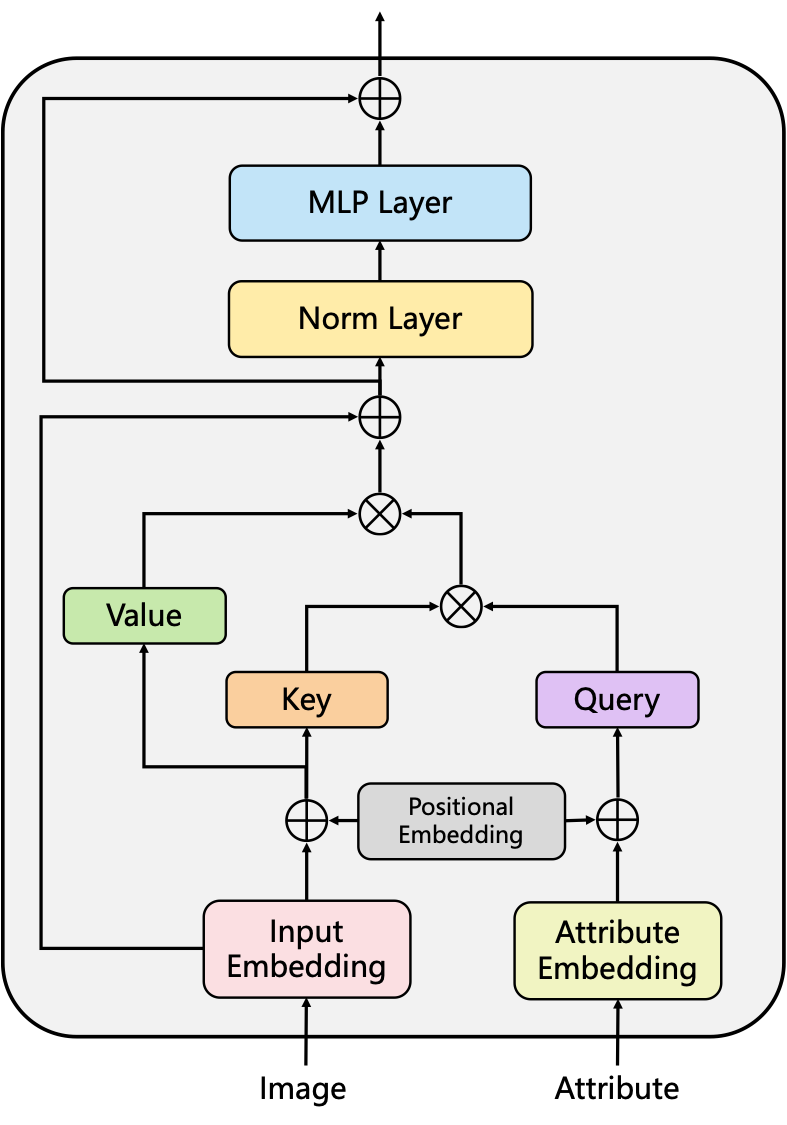}
  \vspace{-20pt}
  \caption{Illustration of the proposed fair identity attention (FIA) Block. This block maps demographic attributes onto attention query vectors, thereby dynamically extracting relevant salient features associated with the current attributes for downstream tasks.}
  \label{fig:fairAtt}
\end{wrapfigure}
based mechanism designed to discern and utilize features from images that are pertinent to the downstream tasks, such as segmentation and classification, while also taking into account the demographic attributes. 

More formally, we seek to optimize a model \( f_\theta : \mathbb{R}^{H \times W} \times \mathbb{R}^{d} \xrightarrow[]{\theta} \mathcal{Y} \), where the outcome space \( \mathcal{Y} \) encompasses both \( \mathbb{R}^K \) for classification or \( \mathbb{R}^{K \times H \times W} \) for segmentation, reflecting the objectives of categorizing images into \( K \) classes or labeling each pixel in an image with one of these \( K \) classes, respectively.

Fig.~\ref{fig:fairAtt} illustrates the architecture of the proposed fair identity attention module. The module initially processes the input image and attribute labels through an embedding layer to obtain the input embedding \( E_i \) and attribute embedding \( E_a \). These embeddings are then added to the position embedding \( E_p \). The sum of \( E_i \) and \( E_p \) is utilized to generate the attention keys \( K_{op} \) and values \( V_{op} \), while the sum of \( E_a \) and \( E_p \) is used to produce the attention query \( Q_g \). The specific formulas are as follows:

\begin{equation}
K = \text{Linear}(E_i + E_p) ,  \quad V = \text{Linear}(E_i + E_p), \quad  Q_{a} = \text{Linear}(E_a + E_p).
\end{equation}
By performing the dot product of the query and keys, we extract a similarity matrix relevant to the current attribute. This matrix is then used in a dot product with the values, extracting features that are significant for each attribute with respect to downstream tasks. This process is expressed by the following equation:
\begin{equation}
FairAttention(Q_{a}, K, V) = \text{softmax}\left(\frac{Q_{a} \cdot K_{^T}}{\sqrt{D}}\right) \cdot V,
\end{equation}
where $D$ is a scaling factor to avoid overly large values in the softmax function. Subsequently, a residual connection adds \( E_i \) to the output of the attention to maintain the integrity of the input information. Finally, a normalization layer and a multi-layer perceptron (MLP) layer are used to further extract features. After performing another residual operation on the output of these two layers, we obtain the final output of the fairness attention module.

The fair identity attention mechanism serves as a powerful and versatile tool designed to enhance model performance while addressing fairness concerns. By explicitly considering demographic attributes such as gender, race, or ethnicity, it ensures that the learned representations do not inadvertently amplify existing biases present in the data. Its architecture allows it to be seamlessly integrated as a plug-in component into any existing network. This modular nature empowers researchers and practitioners to retrofit fair identity attention into their models without the need for extensive modifications to the underlying architecture. Consequently, the fair identity attention module not only contributes to the improvement of model accuracy and fairness in segmentation and classification tasks but also promotes ethical AI practices by facilitating more equitable treatment of diverse groups within datasets.


\section{Experiments}

\subsection{Algorithmic Fairness Across Domain Shifts}
In our experiments, we first analyze fairness in the context of domain shifts within the Cup-Disc Segmentation task. Cup-Disc Segmentation refers to the process of accurately delineating the optic cup and disc in fundus images, which is essential for calculating the cup-to-disc ratio (CDR) — a critical parameter for assessing the progression and risk of glaucoma. This task is pivotal in the field of medical imaging, particularly within the context of diagnosing and managing eye diseases such as glaucoma. Since the cup is a substantial subarea of the disc, we reformulate the segmentation task into cup and rim (the tissue area between the cup and disc border) segmentation to avoid misrepresented performance due to the large overlapped area between the cup and disc.

We investigate the performance of fairness across three different demographics, including gender, race, and ethnicity, across two distinct domains: En face fundus images derived from optical coherence tomography scans and SLO (Scanning Laser Ophthalmoscopy) fundus images. In the subsequent experiments, we selected en face fundus images as the source domain and SLO fundus images as the target domain. The rationale is that en face fundus images are routinely acquired in specialized ophthalmic care settings in comparison to SLO fundus images, leading to significantly greater availability of data. Therefore, we have chosen to position en face fundus images as the source domain, with SLO fundus images as the target domain. For classification, we utilize fundus images from those two domains as source and target domains, categorized into two classes: normal and glaucoma.

\subsubsection{Evaluation Metrics:} 
For evaluating model performance, the Dice coefficient and the intersection over union (IoU) metrics are employed for segmentation assessments, and the area under the receiver operating characteristic curve (AUC) is utilized for classification tasks.
Those traditional metrics for segmentation and classification, while indicative of model performance, do not inherently account for fairness across demographic identity groups. Inspired by~\cite{luo2023glau_fair,tian2024fairseg}, to address the potential tradeoff between model performance and fairness in medical imaging, we use novel equity-scaled performance (ESP) metrics to assess both performance and fairness for both segmentation and classification tasks. 

\begin{table*}[!t]
    \centering
    \caption{Optic cup and rim segmentation performance on the \textbf{FairDomain-Segmentation} dataset using different Domain Adaptation (DA) and Domain Generalization (DG) methods with \textbf{gender} as the demographic attribute.
    }
    \vspace{-10pt}
    \label{tbl:seg_gender}
    \adjustbox{max width=0.95\textwidth}{
    \begin{tabular}{c c L{20ex} C{12ex} C{10ex} C{10ex} C{10ex} C{10ex} C{10ex} C{10ex} C{10ex} C{10ex}}
            \toprule
        & & & \textbf{Overall} & \textbf{Overall} & \textbf{Overall} & \textbf{Overall} & \textbf{Male} & \textbf{Female} & \textbf{Male} & \textbf{Female}\\
        & & \textbf{Method} & \textbf{ES-Dice$\uparrow$} & \textbf{Dice$\uparrow$} & \textbf{ES-IoU$\uparrow$} & \textbf{IoU$\uparrow$} & \textbf{Dice$\uparrow$} & \textbf{Dice$\uparrow$} & \textbf{IoU$\uparrow$} & \textbf{IoU$\uparrow$}\\
        \midrule
        
        \multirow{4}{*}{\textbf{}}
        & \multirow{2}{*}{\rotatebox[origin=c]{90}{Cup}}
        & Baseline (Source) & 0.885 & 0.888 & 0.806 & 0.808 & 0.886 & 0.889 & 0.807 & 0.810\\ 
        && Baseline (Target) & 0.688 & 0.700 & 0.535 & 0.555 & 0.693 & 0.711 & 0.557 & 0.574\\
        \cmidrule{2-11}
        & \multirow{2}{*}{\rotatebox[origin=c]{90}{Rim}}
        & Baseline (Source) & 0.854 & 0.861 & 0.753 & 0.762 & 0.864 & 0.856 & 0.767 & 0.755\\ 
        && Baseline (Target) & 0.485 & 0.495 & 0.336 & 0.342 & 0.486 & 0.507 & 0.334 & 0.353\\ 
        \midrule
        \midrule
        \multirow{8}{*}{\textbf{DA}}
        & \multirow{4}{*}{\rotatebox[origin=c]{90}{Cup}}
        & PixMatch~\cite{melas2021pixmatch} & 0.768 & 0.775 & 0.650 & 0.660 & 0.772 & 0.769 & 0.645 & 0.660 \\
        & & CBST~\cite{zou2018unsupervised} & 0.791 & 0.797 & 0.682 & 0.686 & 0.794 & 0.802 & 0.684 & 0.690\\
        & & DAFormer~\cite{hoyer2022daformer} & 0.781 & 0.785 & 0.676 & 0.680 & 0.783 & 0.789 & 0.678 & 0.684\\
        & & \textbf{DAFormer-FIA} & \textbf{0.802} & \textbf{0.810} & \textbf{0.692} & \textbf{0.700} & \textbf{0.806} & \textbf{0.816} & \textbf{0.695} & \textbf{0.706}\\
        \cmidrule{2-11}
        & \multirow{4}{*}{\rotatebox[origin=c]{90}{Rim}}
        &  PixMatch~\cite{melas2021pixmatch} & 0.660 & 0.673 & 0.519 & 0.523 & 0.669 & 0.688 & 0.519 & 0.528\\
        & & CBST~\cite{zou2018unsupervised} & \textbf{0.693} & \textbf{0.696} & \textbf{0.541} & \textbf{0.544} & \textbf{0.694} & \textbf{0.698} & \textbf{0.542} & \textbf{0.548}\\
        & & DAFormer~\cite{hoyer2022daformer} & 0.344 & 0.345 & 0.212 & 0.213 & 0.344 & 0.347 & 0.212 & 0.214\\
        & & \textbf{DAFormer-FIA} & 0.528 & 0.531 & 0.367 & 0.369 & 0.533 & 0.528 & 0.372 & 0.366\\
        \midrule
        \midrule
        \multirow{8}{*}{\textbf{DG}}
        & \multirow{4}{*}{\rotatebox[origin=c]{90}{Cup}}
        & GIN+IPA~\cite{ouyang2022causality} & 0.741 & 0.750 & 0.590 & 0.594 & 0.740 & 0.752 & 0.590 & 0.597\\
        & & AADG~\cite{lyu2022aadg} & 0.711 & 0.715 & 0.595 & 0.598 & 0.713 & 0.717 & 0.596 & 0.601\\
        & & DAFormer~\cite{hoyer2022daformer} & 0.787 & 0.795 & 0.677 & 0.684 & 0.791 & 0.802 & 0.680 & 0.691\\
        & & \textbf{DAFormer-FIA} & \textbf{0.816} & \textbf{0.820} & \textbf{0.712} & \textbf{0.716} & \textbf{0.818} & \textbf{0.823} & \textbf{0.714} & \textbf{0.719}\\
        \cmidrule{2-11}
        & \multirow{4}{*}{\rotatebox[origin=c]{90}{Rim}}
        & GIN+IPA~\cite{ouyang2022causality} & 0.550 & 0.555 & 0.397 & 0.402 & 0.551 & 0.560 & 0.394 & 0.408\\
        & & AADG~\cite{lyu2022aadg} & 0.621 & 0.621 & 0.465 & 0.466 & 0.621 & 0.621 & 0.465 & 0.467\\
        & & DAFormer~\cite{hoyer2022daformer} & 0.659 & 0.669 & 0.507 & 0.516 & 0.663 & 0.677 & 0.509 & 0.526\\
        & & \textbf{DAFormer-FIA} & \textbf{0.704} & \textbf{0.705} & \textbf{0.557} & \textbf{0.559} & \textbf{0.704} & \textbf{0.706} & \textbf{0.558} & \textbf{0.561}\\
        \bottomrule
    \end{tabular}
    }
\end{table*}

\begin{table*}[!t]
    \centering
    \caption{Optic cup and rim segmentation performance on the \textbf{FairDomain-Segmentation} dataset using different Domain Adaptation (DA) and Domain Generalization (DG) methods with \textbf{race} as the demographic attribute.}
    \label{tbl:seg_race}
    \vspace{-10pt}
    \adjustbox{max width=0.95\textwidth}{
    \begin{tabular}{c c L{20ex} C{12ex} C{10ex} C{10ex} C{10ex} C{10ex} C{10ex} C{10ex} C{10ex} C{10ex} C{10ex}}
            \toprule
        & & & \textbf{Overall} & \textbf{Overall} & \textbf{Overall} & \textbf{Overall} & \textbf{Asian} & \textbf{Black} & \textbf{White} & \textbf{Asian} & \textbf{Black} & \textbf{White} \\
        & & \textbf{Method} & \textbf{ES-Dice$\uparrow$} & \textbf{Dice$\uparrow$} & \textbf{ES-IoU$\uparrow$} & \textbf{IoU$\uparrow$} & \textbf{Dice$\uparrow$} & \textbf{Dice$\uparrow$} & \textbf{Dice$\uparrow$} & \textbf{IoU$\uparrow$} & \textbf{IoU$\uparrow$} & \textbf{IoU$\uparrow$}\\
        \midrule
        \multirow{4}{*}{\textbf{}}
        & \multirow{2}{*}{\rotatebox[origin=c]{90}{Cup}}
        &  Baseline (Source)  & 0.868 & 0.888 & 0.784 & 0.808 & 0.881 & 0.901 & 0.886 & 0.802 & 0.830 & 0.805\\
        & & Baseline (Target) & 0.673 & 0.700 & 0.535 & 0.564 & 0.703 & 0.731 & 0.694 & 0.567 & 0.606 & 0.555\\ 
        \cmidrule{3-13}
        & \multirow{2}{*}{\rotatebox[origin=c]{90}{Rim}}
        & Baseline (Source)  & 0.820 & 0.861 & 0.716 & 0.762 & 0.841 & 0.838 & 0.868 & 0.737 & 0.731 & 0.771\\ 
        & & Baseline (Target) & 0.481 & 0.495 & 0.334 & 0.342 & 0.496 & 0.472 & 0.499 & 0.347 & 0.326 & 0.345\\
        \midrule
        \midrule
        \multirow{8}{*}{\textbf{DA}}
        & \multirow{4}{*}{\rotatebox[origin=c]{90}{Cup}}
       
        & PixMatch~\cite{melas2021pixmatch} & 0.739 & 0.775 & 0.614 & 0.660 & 0.752 & 0.793 & 0.783 & 0.633 & 0.693 & 0.673\\
        & & CBST~\cite{zou2018unsupervised} & 0.770 & 0.791 & 0.662 & 0.679 & 0.773 & 0.785 & 0.794 & 0.657 & 0.680 & 0.681\\
        & & DAFormer~\cite{hoyer2022daformer} & 0.785 & 0.804 & 0.669 & 0.691 & 0.794 & 0.816 & 0.802 & 0.682 & 0.712 & 0.688\\
        & & \textbf{DAFormer-FIA} & \textbf{0.796} & \textbf{0.810} & \textbf{0.682} & \textbf{0.700} & \textbf{0.804} & \textbf{0.821} & \textbf{0.809} & \textbf{0.694} & \textbf{0.717} & \textbf{0.697}\\
        \cmidrule{2-13}
        & \multirow{4}{*}{\rotatebox[origin=c]{90}{Rim}}
        & PixMatch~\cite{melas2021pixmatch} & \textbf{0.627} & 0.673 & 0.493 & 0.523 & \textbf{0.653} & 0.640 & 0.693 & \textbf{0.522} & 0.498 & 0.559\\
        & & CBST~\cite{zou2018unsupervised}  & \textbf{0.627} & \textbf{0.702} & \textbf{0.496} & \textbf{0.556} & 0.650 & \textbf{0.650} & \textbf{0.719} & 0.502 & \textbf{0.506} & \textbf{0.573}\\
        & & DAFormer~\cite{hoyer2022daformer} & 0.326 & 0.332 & 0.199 & 0.202 & 0.334 & 0.345 & 0.329 & 0.204 & 0.212 & 0.199\\
        & & \textbf{DAFormer-FIA} & 0.521 & 0.531 & 0.363 & 0.369 & 0.522 & 0.539 & 0.530 & 0.364 & 0.380 & 0.368\\
        \midrule
        \midrule
        \multirow{10}{*}{\textbf{DG}}
        & \multirow{4}{*}{\rotatebox[origin=c]{90}{Cup}} 
        & GIN+IPA~\cite{ouyang2022causality} & 0.714 & 0.750 & 0.568 & 0.594 & 0.732 & 0.762 & 0.730 & 0.590 & 0.635 & 0.595\\
        & & AADG~\cite{lyu2022aadg} & 0.694 & 0.715 & 0.586 & 0.598 & 0.723 & 0.696 & 0.717 & 0.607 & 0.588 & 0.600\\
        & & DAFormer~\cite{hoyer2022daformer} & 0.769 & 0.798 & 0.673 & 0.688 & 0.771 & 0.809 & 0.799 & 0.662 & 0.706 & 0.687\\
        & & \textbf{DAFormer-FIA} & \textbf{0.787} & \textbf{0.810} & \textbf{0.675} & \textbf{0.701} & \textbf{0.790} & \textbf{0.820} & \textbf{0.811} & \textbf{0.678} & \textbf{0.717} & \textbf{0.700}\\
        \cmidrule{2-13}
        & \multirow{4}{*}{\rotatebox[origin=c]{90}{Rim}}
        & GIN+IPA~\cite{ouyang2022causality} & 0.527 & 0.555 & 0.383 & 0.402 & 0.538 & 0.542 & 0.577 & 0.404 & 0.378 & 0.424\\
        & & AADG~\cite{lyu2022aadg} & 0.578 & 0.621 & 0.436 & 0.466 & 0.597 & 0.581 & 0.632 & 0.443 & 0.430 & 0.476\\
        & & DAFormer~\cite{hoyer2022daformer} & 0.616 & 0.676 & 0.481 & 0.527 & 0.638 & 0.631 & 0.689 & 0.490 & 0.481 & 0.540\\
        & & \textbf{DAFormer-FIA} & \textbf{0.636} & \textbf{0.689} & \textbf{0.499} & \textbf{0.540} & \textbf{0.655} & \textbf{0.650} & \textbf{0.701} & \textbf{0.509} & \textbf{0.500} & \textbf{0.552}\\
        \bottomrule
    \end{tabular}}
    \vspace{-20pt}
\end{table*}
Let $\mathcal{M} \in \{\text{Dice}, \text{IoU}, \text{AUC}, \ldots\}$ signify a generic performance metric applicable to either segmentation or classification. Traditional evaluation usually takes a set of triplets $(z', a, y)$ as input to produce the metric score $\mathcal{M}(\{(z', y)\})$, which typically disregards demographic identity attributes, thereby missing critical fairness assessment. To incorporate fairness, we first compute a performance discrepancy $\Delta$, defined as the aggregate deviation of each demographic group’s metric from the overall performance, expressed as:
\begin{equation}
    \Delta = \sum_{A \in \mathcal{A}} |\mathcal{M}(\{(z', y)\}) - \mathcal{M}(\{(z', a, y)|a=A\})|,
\end{equation}
where $\Delta$ approaches zero when performance equity across groups is achieved, reflecting minimal disparity.
The ESP metric can then be formulated as follows:
\begin{equation}
    \text{ESP} = \frac{\mathcal{M}(\{(z', y)\})}{1+\Delta},
\end{equation}
which ensures a balanced evaluation of model accuracy and fairness. This metric, \(\text{ESP}\), aligns with \(\mathcal{M}\) when \(\Delta\) is minimized, indicating equitable performance across demographics. Conversely, an increased \(\Delta\) denotes significant disparities between different demographic groups, lowering the ESP score with a larger penalty to indicate that the models obtain less fairness. This unified metric facilitates a comprehensive assessment of deep learning models, emphasizing not only their accuracy (as measured by segmentation and classification metrics such as Dice, IoU and AUC) but also their fairness across different demographic groups.

\subsubsection{Cup-Rim Segmentation Results under Domain Shifts} 

\paragraph{Baselines:} We first utilized TransUNet as the baseline model to perform training on the source domain. After the model training phase, we directly evaluated the model's performance on the target domain without any domain adaptation or generalization strategies. The results of this baseline model on source and target domains are detailed in the first four rows of Table~\ref{tbl:seg_gender}, Tables~\ref{tbl:seg_race}, and Table~\ref{tbl:seg_ethnicity}.  It can be observed that a significant decrease in segmentation accuracy for both the optic cup and rim when applying a model trained on the source domain directly to the target domain. This performance decline highlights the critical issue of domain shift in medical imaging segmentation tasks, underscoring the challenge of transferring learned models between differing imaging domains without appropriate adaptation or generalization strategies.

\begin{table*}[!t]
    \centering
    \caption{Optic Cup and Rim segmentation performance on the \textbf{FairDomain-Segmentation} dataset using different Domain Adaptation (DA) and Domain Generalization (DG) methods with \textbf{ethnicity} as the demographic attribute.}
    \label{tbl:seg_ethnicity}
    \vspace{-10pt}
    \adjustbox{max width=0.95\textwidth}{
    \begin{tabular}{c c L{20ex} C{12ex} C{10ex} C{10ex} C{10ex} C{10ex} C{18ex} C{10ex} C{18ex}}
            \toprule
        & & & \textbf{Overall} & \textbf{Overall} & \textbf{Overall} & \textbf{Overall} & \textbf{Hispanic} & \textbf{Non-Hispanic} & \textbf{Hispanic} & \textbf{Non-Hispanic}\\
        & & \textbf{Method} & \textbf{ES-Dice$\uparrow$} & \textbf{Dice$\uparrow$} & \textbf{ES-IoU$\uparrow$} & \textbf{IoU$\uparrow$} & \textbf{Dice$\uparrow$} & \textbf{Dice$\uparrow$} & \textbf{IoU$\uparrow$} & \textbf{IoU$\uparrow$}\\
        \midrule
        \multirow{4}{*}{\textbf{}}
        & \multirow{2}{*}{\rotatebox[origin=c]{90}{Cup}}
        & Baseline (Source) & 0.871 & 0.888 & 0.784 & 0.808 & 0.887 & 0.906 & 0.807 & 0.839\\ 
        & & Baseline (Target) & 0.684 & 0.700 & 0.551 & 0.555 & 0.699 & 0.722 & 0.563 & 0.587\\ 
        \cmidrule{2-11}
        & \multirow{2}{*}{\rotatebox[origin=c]{90}{Rim}}
        & Baseline (Source) & 0.845 & 0.861 & 0.743 & 0.762 & 0.860 & 0.879 & 0.761 & 0.786\\ 
        & & Baseline (Target) &  0.489 & 0.495 & 0.340 & 0.342 & 0.495 & 0.506 & 0.342 & 0.348\\ 
        \midrule
        \midrule
        \multirow{8}{*}{\textbf{DA}}
        & \multirow{4}{*}{\rotatebox[origin=c]{90}{Cup}}
        & PixMatch~\cite{melas2021pixmatch} & 0.740 & 0.775 & 0.628 & 0.660 & 0.763 & 0.811 & 0.649 & 0.699\\
        & & CBST~\cite{melas2021pixmatch} & 0.768 & 0.797 & 0.655 & 0.686 & 0.794 & 0.833 & 0.683 & 0.730\\
        & & DAFormer~\cite{hoyer2022daformer} & 0.773 & 0.796 & 0.654 & 0.680 & 0.794 & 0.824 & 0.677 & 0.717\\
        & & \textbf{DAFormer-FIA} & \textbf{0.790} & \textbf{0.810} & \textbf{0.674} & \textbf{0.700} & \textbf{0.808} & \textbf{0.834} & \textbf{0.697} & \textbf{0.734}\\
        \cmidrule{2-11}
        & \multirow{4}{*}{\rotatebox[origin=c]{90}{Rim}}
        & PixMatch~\cite{melas2021pixmatch} & 0.643 & 0.673 & 0.499 & 0.523 & 0.673 & 0.719 & 0.513 & 0.562\\
        & & CBST~\cite{zou2018unsupervised} & \textbf{0.675} & \textbf{0.696} & \textbf{0.528} & \textbf{0.544} & \textbf{0.693} & \textbf{0.723} & \textbf{0.542} & \textbf{0.573} \\
        & & DAFormer~\cite{hoyer2022daformer} & 0.405 & 0.418 & 0.264 & 0.270 & 0.416 & 0.448 & 0.269 & 0.295\\
        & & \textbf{DAFormer-FIA} & 0.516 & 0.531 & 0.359 & 0.369 & 0.529 & 0.558 & 0.368 & 0.395\\
        \midrule
        \midrule
        \multirow{8}{*}{\textbf{DG}}
        & \multirow{4}{*}{\rotatebox[origin=c]{90}{Cup}}
        & GIN+IPA~\cite{ouyang2022causality} & 0.714 & 0.750 & 0.557 & 0.594 & 0.710 & 0.760 & 0.540 & 0.607\\
        & & AADG~\cite{lyu2022aadg} & 0.705 & 0.715 & 0.594 & 0.598 & 0.715 & 0.700 & 0.597 & 0.605\\
        & & DAFormer~\cite{hoyer2022daformer} & 0.746 & 0.746 & 0.630 & 0.640 & 0.746 & 0.746 & 0.640 & 0.656\\
        & & \textbf{DAFormer-FIA} & \textbf{0.785} & \textbf{0.810} & \textbf{0.672} & \textbf{0.701} & \textbf{0.809} & \textbf{0.842} & \textbf{0.699} & \textbf{0.743} \\
        \cmidrule{2-11}
        & \multirow{4}{*}{\rotatebox[origin=c]{90}{Rim}}
        & GIN+IPA~\cite{ouyang2022causality} & 0.536 & 0.555 & 0.391 & 0.402 & 0.537 & 0.572 & 0.398 & 0.426\\
        & & AADG~\cite{lyu2022aadg} & 0.603 & 0.621 & 0.453 & 0.466 & 0.620 & 0.649 & 0.465 & 0.494\\
        & & DAFormer~\cite{hoyer2022daformer} & 0.630 & 0.641 & 0.491 & 0.499 & 0.642 & 0.624 & 0.499 & 0.485\\
        & & \textbf{DAFormer-FIA} & \textbf{0.676} & \textbf{0.689} & \textbf{0.531}& \textbf{0.540} & \textbf{0.688} & \textbf{0.708} & \textbf{0.540} & \textbf{0.557}\\
        \bottomrule
    \end{tabular}}
\end{table*}

\paragraph{Domain Adaptation for Segmentation:} To assess the performance of existing domain adaptation methods, we selected three state-of-the-art models as baseline methods for domain adaptation: \textbf{PixMatch}\cite{melas2021pixmatch}, \textbf{CBST}\cite{zou2018unsupervised}, and \textbf{DAFormer}~\cite{hoyer2022daformer}. Given that DAFormer has demonstrated superior performance over the other two methods in previous tasks, and considering its applicability to both domain adaptation and domain generalization tasks, we have integrated our proposed fair identity attention mechanism into DAFormer, denoted as \textbf{DAFormer-FIA} to evaluate the effectiveness of our model. Table~\ref{tbl:seg_gender}, Tables~\ref{tbl:seg_race}, and Table~\ref{tbl:seg_ethnicity} present the domain adaptation results of these four models across three demographic attributes: gender, race, and ethnicity. It is noteworthy that our DAFormer-FIA achieves improvements in the segmentation of both the cup and rim across all attributes compared to the baseline DAFormer. Specifically, our model improved the ES-Dice for cup segmentation from 0.781 to 0.802 and the ES-Dice for rim segmentation from 0.344 to 0.528 in the gender attribute. For race attribute, it increased the ES-Dice for cup segmentation from 0.785 to 0.796, and for rim segmentation, from 0.326 to 0.521. In terms of ethnicity attribute, the ES-Dice for cup segmentation was enhanced from 0.773 to 0.790, and the ES-Dice for rim segmentation was elevated from 0.405 to 0.516. Concurrently, for cup segmentation, our model achieved the highest ES-Dice scores of 0.796 for gender, 0.796 for race, and 0.790 for ethnicity. Similarly, it secured the best ES-IOU scores of 0.692 for gender, 0.682 for race, and 0.674 for ethnicity.  Regarding specific demographic attributes, DAFormer-FIA increased the Dice for rim segmentation in the male category within gender from 0.344 to 0.533, marking an improvement of approximately 54.94\%. In the white category within race, it improved from 0.329 to 0.530, a 61.09\% increase. Lastly, in the Hispanic category within ethnicity, the Dice was enhanced from 0.416 to 0.529, translating to a 27.16\% uplift. These substantial percentage improvements underscore the effectiveness of our model across diverse demographic attributes. This consistent enhancement across various demographic attributes highlights the effectiveness of integrating fair identity attention into domain adaptation tasks, particularly in augmenting the segmentation accuracy of critical ocular structures.

\paragraph{Domain Generalization for Segmentation:}
To validate the performance of existing domain generalization methods, we selected three state-of-the-art models as baselines for domain generalization: GIN+IPA~\cite{ouyang2022causality}, AADG~\cite{lyu2022aadg}, and DAForm-er~\cite{hoyer2022daformer}. Similarly, we integrated our proposed fair identity attention mechanism into DAFormer to assess the effectiveness of our model. Tables~\ref{tbl:seg_race}, Table~\ref{tbl:seg_gender}, and Table~\ref{tbl:seg_ethnicity} outline the domain generalization results of these four models across three demographic attributes: race, gender, and ethnicity. It can be found from these tables that our DAFormer-FIA not only enhances the segmentation of both the cup and rim across all three attributes over the standard DAFormer but also achieves the highest segmentation performance for both the cup and rim across all three attributes. Specifically, our model improved the ES-Dice score for cup segmentation from 0.787 to 0.816 and for rim segmentation from 0.659 to 0.704 in the gender attribute. Within the race attribute, there were increments in the ES-Dice scores for cup and rim segmentations, from 0.769 to 0.787 and 0.616 to 0.636, respectively. Regarding ethnicity, the model’s cup segmentation ES-Dice improved from 0.746 to 0.785, and rim segmentation from 0.630 to 0.676. Regarding specific demographic attributes, DAFormer-FIA increased the Dice score for rim segmentation in the male category within gender from 0.663 to 0.704. In the black category within race, the IoU improved from 0.329 to 0.530. In the Non-Hispanic category within ethnicity, the Dice score was enhanced from 0.746 to 0.842. This indicates the significant impact of incorporating fair identity attention into domain generalization tasks, demonstrating substantial improvements in segmentation accuracy across diverse demographic attributes.

\subsection{Glaucoma Classification Results under Domain Shifts}
\paragraph{Baselines:} As shown in Table~\ref{tbl:cls_overall}, the basic overall AUC of using en face fundus images (i.e. source domain) to predict glaucoma is 0.803, while the ES-AUCs for gender, race, and ethnicity are 0.795, 0.730, and 0.744, respectively. There are significant AUC performance disparities for racial subgroups with Blacks are 0.085 lower than Whites and 0.058 lower than Asians. Similarly, Hispanics achieve significantly lower AUC than Non-Hispanics. However, if using slo fundus to predict glaucoma with the pretrained source-domain model, both overall AUC and group AUCs have dramatic drops for gender, racial, and ethnic groups. As shown in Fig. \ref{fig:gd}, the group disparities generally become worse while transforming the source domain to the target domain for both segmentation and classification tasks. This suggests that source and target domain fundus images present signific-antly different semantic distributions. It is necessary to adopt specific learning module to minimize such a domain shift and meanwhile reduce group performance disparities.

\begin{table*}[!t]
    \centering
    \caption{Classification performance on the \textbf{FairDomain-Classification} dataset using different Domain Adaptation (DA) and Domain Generalization (DG) methods with \textbf{gender}, \textbf{race}, and \textbf{ethnicity} as the demographic attribute.} 
    \label{tbl:cls_overall}

    \vspace{-10pt}
    \adjustbox{max width=1\textwidth}{
    \begin{tabular}{c L{20ex} C{12ex} C{10ex} C{10ex}  C{10ex} C{12ex} C{10ex} C{10ex} C{10ex} C{10ex} C{12ex} C{10ex} C{18ex}  C{10ex}}
        \toprule
        & & \textbf{Gender Overall} & \textbf{Gender Overall} & \textbf{Male Group} & \textbf{Female Group} & \textbf{Race Overall} & \textbf{Race Overall} & \textbf{Asian Group} & \textbf{Black Group} & \textbf{White Group} & \textbf{Ethnicity Overall} & \textbf{Ethnicity Overall} & \textbf{Non-Hispanic Group} & \textbf{Hispanic Group}\\
        & \textbf{Method} & \textbf{ES-AUC}$\uparrow$ & \textbf{AUC}$\uparrow$ & \textbf{AUC}$\uparrow$  & \textbf{AUC}$\uparrow$ & \textbf{ES-AUC}$\uparrow$ & \textbf{AUC}$\uparrow$ & \textbf{AUC}$\uparrow$ & \textbf{AUC}$\uparrow$ & \textbf{AUC}$\uparrow$ & \textbf{ES-AUC}$\uparrow$ & \textbf{AUC}$\uparrow$ & \textbf{AUC}$\uparrow$  & \textbf{AUC}$\uparrow$\\
        \midrule
        & Baseline (Source) & 0.795 & 0.803 & 0.806 & 0.796 & 0.730 & 0.803 & 0.788 & 0.730 & 0.815 & 0.744 & 0.803 & 0.806 & 0.727\\
        & Baseline (Target) & 0.536 & 0.537 & 0.536 & 0.538 & 0.508 & 0.537 & 0.555 & 0.511 & 0.524 & 0.517 & 0.537 & 0.539 & 0.501\\ \midrule
        \multirow{3}{*}{\rotatebox[origin=c]{90}{\textbf{DA}}}
        & CGDM~\cite{du2021cross} & 0.611 & 0.618 & 0.613 & 0.625& 0.556 & 0.618 & 0.662 & 0.604 & \textbf{0.671} & 0.579 & 0.618 & 0.620 & 0.554\\
        & CDTrans~\cite{xu2021cdtrans} & 0.631 & 0.633 & 0.634 & 0.631 & 0.603 & 0.633 & 0.658 & 0.614 & 0.628 & 0.600 & 0.633 & 0.635 & 0.579\\
        & \textbf{CDTrans-FIA} & \textbf{0.633} & \textbf{0.635} & \textbf{0.636} & \textbf{0.633} & \textbf{0.606} & \textbf{0.636} & \textbf{0.663} & \textbf{0.619} & 0.631 & \textbf{0.607} & \textbf{0.636} & \textbf{0.638} & \textbf{0.589}\\
        \midrule \midrule
        \multirow{4}{*}{\rotatebox[origin=c]{90}{\textbf{DG}}}
        & GroupDro~\cite{sagawa2019distributionally}  & 0.513 & 0.530 & 0.538 & 0.506 & 0.481 & 0.530 & 0.578 & 0.512 & 0.495 & 0.501 & 0.530 & 0.532 & 0.476\\
        & IRM~\cite{arjovsky2019invariant}  & 0.667 & 0.672 & 0.671 & \textbf{0.678} & 0.666 & 0.672 & 0.668 & 0.677 & 0.672 & 0.601 & 0.672 & 0.677 & 0.558\\
        & G2DM~\cite{albuquerque2019generalizing} & 0.610 & 0.616 & 0.618 & 0.609 & 0.593 & 0.616 & 0.606 & 0.611 & 0.592 & 0.607 & 0.616 & 0.616 & \textbf{0.602}\\
        & \textbf{IRM-FIA} & \textbf{0.670} & \textbf{0.676} & \textbf{0.678} & 0.670 & \textbf{0.671} & \textbf{0.696} & \textbf{0.679} & \textbf{0.693} & \textbf{0.680} & \textbf{0.614} & \textbf{0.674} & \textbf{0.680} & 0.580\\
        \bottomrule
    \end{tabular}}
    \vspace{-10pt}
\end{table*}

\paragraph{Domain Adaptation for Classification:}

In parallel with our approach for segmentation, we explored two leading domain adaptation (DA) methods for classification tasks: \textbf{CGDM}~\cite{du2021cross} and \textbf{CDTrans}~\cite{xu2021cdtrans}, selecting the latter, \textbf{CDTrans}, for enhancement with our novel fair identity attention module, yielding \textbf{CDTrans-FIA}. This integration aims to bolster fairness across demographics under DA. Performances across gender, race, and ethnicity for these models are showcased in Table~\ref{tbl:cls_overall}. Our proposed CDTrans-FIA stands out by securing substantial improvements across all demographic attributes. It records the highest ES-Dice scores 0.633 for gender, 0.606 for race, and 0.607 for ethnicity, demonstrating the efficacy of our proposed fair identity attention in domain adaptation, especially for enhancing glaucoma classification algorithmic fairness. Additionally, the proposed method (Table~\ref{tbl:cls_overall}) significantly uplifts accuracy for minority groups; it boosts performance for blacks by 0.005 to 0.016 AUC scores, and for Hispanics by 0.01 to 0.035 AUC scores, outperforming other DA approaches. Moreover, our model exhibits notable advancements in both accuracy and fairness in the target domain, surpassing the baseline with significant improvements of around 0.1 AUC scores in overall AUC and ES-AUC.

\paragraph{Domain Generalization for Classification:}
Similarly, we selected three state-of-the-art methods including GroupDro \cite{sagawa2019distributionally}, IRM \cite{araslanov2021self}, and G2DM \cite{du2021cross} for the DA-based classification. Given that IRM generally achieves the best overall AUC performances for all three demographic attributes, we select IRM as the backbone architecture to incorporate our proposed fair identity attention module, termed as IRM-FIA.  According to the results presented in Table~\ref{tbl:cls_overall}, IRM-FIA surpasses both GroupDro and G2DM in overall AUC and ES-AUC scores, proving its superior capability in facilitating equitable glaucoma classification across diverse identity groups. In addition, IRM-FIA is generally superior to IRM in both overall AUC and ES-AUC for all three attributes, with the improvement most prominent for race. For instance, as detailed in Table~\ref{tbl:cls_overall}, there was an increase of 0.024 in the overall AUC, with the AUC for Asians, Blacks, and Whites each showing gains of over 0.01 AUC score. This underscores the effectiveness of our fair identity attention module in not only boosting overall classification accuracy but also minimizing disparities among different subgroups.

\section{Conclusion}
To summarize, this paper focuses on addressing fairness in AI, especially in medical AI, which is essential for equitable healthcare. The issue of fairness in domain transfer, due to the fact that clinics may use varied imaging technologies, remains largely unexplored. Our work introduces FairDomain, a comprehensive study on algorithmic fairness in domain transfer tasks including domain adaptation and generalization for both medical segmentation and classification. We propose a novel plug-and-play fair identity attention module that enhances fairness by learning feature relevance through the attention mechanism according to demographic attributes in domain transfer tasks. We also create the first fairness-centric dataset with two paired imaging modalities for the same patient cohort to exclude the confounding impact of demographic distribution variation on model fairness to allow precise assessment of the impact of domain shift on model fairness. Our fair identity attention model can improve existing domain adaptation and generalization methods with better model performance accounted for fairness.

\clearpage  

%
%
\bibliographystyle{splncs04}
\bibliography{main}

\newpage
\appendix
\section{More training details} 
We conducted the experiments on a single A100 GPU with 80GB of memory. For each baseline, we adhered to the training settings specified in the original paper. For the proposed DAFormer-FIA, which is designed for segmentation tasks, we added one FIA layer after the original encoder module at each feature learning stage during the downsampling process. We trained the model using the AdamW optimizer. The encoder was trained with a base learning rate of $6e-5$, and the decoder with $6e-4$. The model was trained with a batch size of 2 for 40k iterations. For CDTrans-FIA in the domain adaptation task, we considered demographic attributes for query in the cross-attention layer of the backbone ViT model and followed the same training parameters in CDTrans. IRM-FIA was designed for the domain generalization task, which incorporated one FIA layer after the feature encoder module of the backbone SWIM model. IRM-FIA followed the default parameters in IRM.

\section{Computational complexity} The FLOPs comparison between IRM+FIA vs IRM is $5.1e11$ vs. $4.9e11$. The training time per epoch and inference time per sample are $157$s vs. $149$s and $0.70$s vs. $0.65$s, respectively. 

\end{document}